\documentclass[twocolumn,secnumarabic,amssymb, aps, prl]{revtex4-1}

\usepackage{graphicx}
\usepackage{dcolumn}
\usepackage{bm}
\usepackage{epsfig}
\usepackage{epstopdf}
\usepackage{ulem}
\usepackage{color}

\begin{document}

\title{Metal-insulator transition in ultrathin LaNiO$_3$ films}

\author{R. Scherwitzl$^1$}
\email{raoul.scherwitzl@unige.ch}
\author{S. Gariglio$^1$}
\author{M. Gabay$^2$}
\author{P. Zubko$^1$}
\author{M. Gibert$^1$}
\author{J.-M. Triscone$^1$}

\affiliation{$^1$DPMC, University of Geneva, 24 Quai Ernest-Ansermet, 1211 Gen\`eve 4, Switzerland}
\affiliation{$^2$Laboratoire de Physique des Solides, Universit\'e Paris-Sud 11, Centre dêOrsay, 91405 Orsay Cedex, France}
\date{\today}%

\begin{abstract}
Transport in ultrathin films of LaNiO$_3$ evolves from a metallic to a strongly localized character as the film's thickness is reduced and the sheet resistance reaches a value close to $h/e^2$, the quantum of resistance in two dimensions. In the intermediate regime, quantum corrections to the Drude low-temperature conductivity are observed; they are accurately described by weak localization theory. Remarkably, the negative magnetoresistance in this regime is isotropic, which points to magnetic scattering associated with the proximity of the system to either a spin glass state or the charge ordered antiferromagnetic state observed in other rare earth nickelates.
\end{abstract}
\maketitle

\begin{figure}[tb]
\includegraphics[width=\columnwidth]{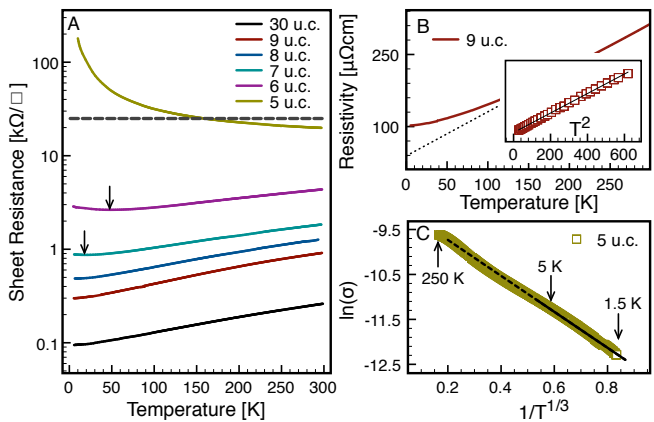}
\caption{A: Sheet resistance versus temperature for different film thicknesses. The dashed line corresponds to $\sim 25\,\rm{k}\Omega$, the quantum of resistance in 2D. Arrows mark the temperatures where upturns in the sheet resistance occur. B: Resistivity versus temperature for a metallic 9 u.c. film. The dotted line is an extrapolation of the high temperature linear behavior. The inset shows the resistivity versus T$^2$ for temperatures below 25~K. C: Logarithm of conductance as a function of $1/T^{1/3}$ for a 5 u.c. film for temperatures from 1.5 K to 250 K. The black line is the linear fit to the data between 1.5 K and 5 K.}
\label{Fig1}
\end{figure}

\begin{figure}[tb]
\includegraphics[width=0.8\columnwidth]{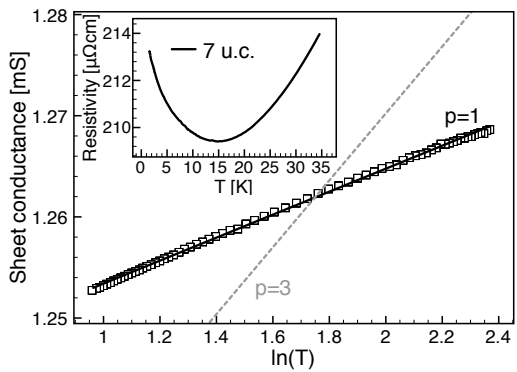}
\caption{Sheet conductance versus the logarithm of temperature for a 7 u.c. film. The linear fit to the data (black line) gives a slope of 1.1$\cdot 10^{-5}$ S $\approx e^2/\pi h$, hence $p=1$. If $p=3$, the data points should follow the grey dotted line. The inset shows the resistivity of the same sample on a linear scale.}
\label{Fig2}
\end{figure}

\begin{figure}[tb]
\includegraphics[width=\columnwidth]{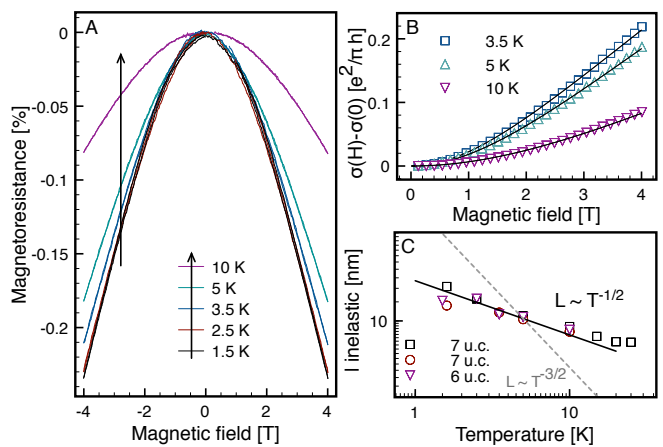}
\caption{A: Perpendicular magnetic field dependence of the MR, $[R(H)-R(0)]/R(0)$, of a 7 u.c. film for several fixed temperatures. B: Magnetoconductance as a function of magnetic field including fits to the data discussed in the text. The number of points and temperature curves were reduced for clarity. C: Temperature behaviour of the inelastic scattering length, extracted from the fits of the magnetoconductance for three different samples showing weak localization. The solid and dotted lines are theoretical curves for the case where $p=1$ and 3 respectively.}
\label{Fig3}
\end{figure}

\begin{figure}[tb]
\includegraphics[width=0.75\columnwidth]{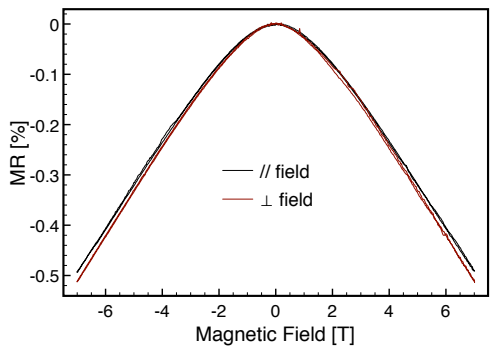}
\caption{Magnetoresistance with the magnetic field applied parallel and perpendicular to the film plane, measured at 1.5 K for a 7 u.c. LNO film}
\label{Fig4}
\end{figure}

Complex oxides are exciting materials in which the interplay between charge, spin, orbital and lattice degrees of freedom leads to a wealth of novel and exotic phenomena.
Among these materials, LaNiO$_3$ (LNO), a metal and paramagnet lacking any ordering phenomena in bulk \cite{RAJEEV:1991p968,SREEDHAR:1992p1956,Xu:1993p1948}, has recently become the subject of intense research. This was mainly triggered  by theoretical work \cite{Chaloupka:2008p77} which suggested the possibility of orbital ordering and high-T$_{\rm{c}}$ superconductivity in superlattices where very thin LNO layers are separated by insulating layers, confining the conduction to two dimensions (2D). These predictions have yet to be confirmed experimentally.\\ 
LNO bulk and thin films have been extensively studied in the past, showing a large effective mass and enhanced Pauli paramagnetism, which has been attributed to strong correlations \cite{RAJEEV:1991p968,SREEDHAR:1992p1956, Ouellette:2010p9027}. This material has also been reported to be sensitive to the degree of disorder. For instance, increasing the amount of oxygen vacancies can lead to weak localization and metal-insulator (MI) transitions, as well as antiferromagnetic ordering and spin glass behaviour \cite{Sanchez:1996p56,Gayathri:1998p1857, Herranz2005}.\\
LNO based heterostructures have been grown and transport measurements have revealed a MI transition as the LNO thickness is reduced to only a few unit cells (u.c.)\cite{May:2009p3363,Son:2010p9026}.  This behaviour is in agreement with the MI transition observed in single ultrathin LNO films \cite{Scherwitzl:2009p6057}. \\
Understanding the nature of transport for this compound in the ultrathin limit appears essential for future investigation of LNO-based heterostructures and serves as the motivation for this work. Upon reduction of the LNO thickness, transport evolves from a metallic to a strongly localized behavior as the room temperature sheet resistance approaches $h/e^2\sim 25\,\rm{k}\Omega$, the quantum of resistance in 2D.\ We find a transition region between the metallic and the strongly localized states where the films show an upturn in the resistivity at low temperatures which can be explained by the theory of weak localization. Surprisingly, the magnetoresistance (MR) is found to be isotropic in this regime. We attribute this to the presence of magnetic scattering and discuss its possible origins.\\
Films were grown on (001) SrTiO$_3$ substrates by off-axis rf magnetron sputtering, leading to fully strained epitaxial c-axis oriented LNO.\ Magneto-transport measurements were carried out between 1.5 K and 300 K in a He flux cryostat equipped with a superconducting magnet.\ More details on the growth, structural characterization as well as sample preparation for transport measurements can be found in Ref.~\cite{Scherwitzl:2009p6057}.\\
Ioffe and Regel pointed out that the simple kinetic theory of conductivity has to break down when the particle wavelength is longer than the mean free path $l$. When $k_Fl$ is close to unity,  $k_F$ being the Fermi wave number, a MI transition is expected. In 2D \cite{LICCIARDELLO:1975p6964}, the sheet resistance, $R_{\rm{sheet}}$, is related to $k_Fl$ by
\begin{equation}
k_Fl= \frac{h/e^2}{R_{\rm{sheet}}}\approx \frac{25\,\rm{k}\Omega/\square}{R_{\rm{sheet}}}.
\end{equation}
When quantum corrections are included in the calculation of transport properties \cite{Lee:1985p948}, the above criterion defines a crossover region between weak and strong localization.
Figure \ref{Fig1}A shows the evolution of sheet resistance versus temperature curves with LNO thickness. The sheet resistance increases continuously as the film thickness is reduced and three different regimes can be distinguished. For low sheet resistance values (below $\sim$ 2 k$\Omega$ at room temperature), the films remain metallic down to the lowest temperature of our experiments. This corresponds to films that are at least 8 u.c. thick.\ Below this thickness and for sheet resistance values less than 10 k$\Omega$, we observe an upturn in the resistivity at around 15 K (7 u.c. film) and 40 K (6 u.c. film). Finally, the curve for 5 u.c. thick LNO has a sheet resistance very close to $h/e^2$ at room temperature, resulting in $k_Fl\sim 1$ and insulating behavior is found at all temperatures.\ The resistance of films with lower thicknesses is so large that it lies outside of the accessible measurement range. These three regimes also pertain to films grown on different substrates, such as LaAlO$_3$, NdGaO$_3$, KTaO$_3$, and MgO (as well as LaAlO$_3$, LSAT  and DyScO$_3$ in Ref. \cite{Son:2010p6999}).\\
We may conclude that the  LNO films undergo a MI transition upon reducing the number of layers and that strong localization effects are seen in the 5 u.c. sample, where $k_Fl\sim 1$ at room temperature.\\
In the metallic regime, the films behave as shown in Figure \ref{Fig1}B.\ The resistivity drops by nearly a factor of three from its room temperature value to a residual resistivity of $\rho_0 \sim 100\,\mu\Omega\rm{cm}$. At high temperatures, the resistivity varies linearly with T, with $\delta\rho/\delta T= 0.9\,\mu\Omega\rm{cmK}^{-1}$. At low-temperatures, below 25 K (see inset), the resistivity shows a quadratic temperature dependence, $\rho=\rho_0+AT^2$ with $A=7.7\cdot 10^{-3}\,\mu\Omega\rm{cmK}^{-2}$. Further characterization can be found in Ref. \cite{Scherwitzl:2009p6057}. This behavior and the characteristic values above are typical of bulk stoichiometric LNO \cite{SREEDHAR:1992p1956}.\\
Attempts to fit the T dependence of the conductivity in the insulating state of 5 u.c. thick films to a simple activated form ($\ln(\sigma)\sim 1/T$) failed, as there is a clear curvature in the Arrhenius plot. For this strongly localized regime, we expect a variable range hopping (VRH) type of conduction. In this case, at low temperatures, electrons hop between localized states and the sheet conductance is given by \cite{BRENIG:1973p7032}
\begin{equation}
\sigma = C \exp(-(T_0/T)^{\alpha}),
\end{equation}
where $T_0$ depends on the density of localized states at the Fermi level and the falloff rate of the wave functions associated with these states. For non-interacting, $d$-dimensional systems, the exponent $\alpha=\frac{1}{d+1}$. Attempts to model the data with a 3D expression ($\alpha=1/4$) were not successful.
Figure \ref{Fig1}C shows the logarithm of the sheet conductance as a function of $1/T^{1/3}$ for a temperature range from 1.5~K to 250~K. Below 5 K, excellent agreement with a linear fit can be observed (with $T_0=4.0$ K and $C=0.13$ mS), consistent with VRH in 2D with $\alpha=1/3$. However, due to the limited temperature range of the fit, it is difficult to distinguish between the exponent $\alpha=1/3$ and $\alpha=1/2$, which is the relevant exponent if Coulomb interactions play a role.
At higher temperatures, the behavior of the resistivity deviates only slightly from the linear fit, indicating that VRH may still be the main conduction mechanism. Indeed, it was argued in Ref.~\cite{EMIN:1974p7033} that VRH can persist at higher temperatures due to polaron assisted hopping.\\
The conductance in the insulating state is consistent with 2D VRH, which also indicates that the thickness driven metal-insulator transition can be associated with a dimensional crossover from a 3D metallic state to a 2D insulating state.\\
As mentioned before, when the film thickness is reduced to 7 and 6 u.c., an upturn in the resistivity appears at low temperature, as indicated by the arrows in Figure \ref{Fig1}A. The inset of Figure \ref{Fig2} shows the low-temperature behavior of the resistivity on a linear scale for a 7 u.c. film.\ 
A plausible interpretation of this phenomenon is weak localization
where quantum interference of electronic waves diffusing around impurities enhances backscattering  and thus leads to a reduction of the conductivity.
In the 2D case \cite{Lee:1985p948}
\begin{equation}
\sigma=\sigma_0 + p\frac{e^2}{\pi h}\ln(T/T_0') \label{Eq1}
\end{equation}
where $\sigma_0$ is the Drude conductivity and $p$ the temperature exponent of the inelastic scattering length $l_{in}\sim T^{-p/2}$. If the main inelastic relaxation mechanism is due to electron-electron collisions, $p=1$, whereas electron-phonon scattering  gives rise to $p=3$ \cite{Lee:1985p948}. $T_0'$ is related to the transport mean free path, which represents the lower length cutoff for diffusive motion.
Figure \ref{Fig2} shows the sheet conductance as a function of $\ln(T)$ for a 7 u.c. film. A linear fit to this curve yields a slope of 1.1$\cdot 10^{-5}$ S $\approx e^2/\pi h$, which is precisely equal to the theoretical prediction given by Equation \ref{Eq1} with $p=1$.\ This points to electron-electron collisions as the main inelastic scattering mechanism. The remaining fitting parameter $\sigma_0-p\frac{e^2}{\pi h}\ln(T_0')$ returns a value of 1.2 mS. For 6 u.c. thick films, the slope is 1.2$\cdot 10^{-5}$ S, again very close to $e^2/\pi h$.\ The dotted line in Figure \ref{Fig2} indicates how the conductance would behave if electron-phonon collisions were dominant.\ 
At this stage, two points should be noted. First, a 2D model was used to describe the low-temperature behavior, which requires that the inelastic scattering length be larger than the thickness of the film. As will be shown later, this constraint indeed appears to be fulfilled in our samples and we have furthermore checked that the data cannot be fitted to a three dimensional model where the conductivity goes as $T^{p/2}$ \cite{Lee:1985p948}. Nevertheless, the validity of the 2D assumption needs further confirmation. Second, Coulomb interactions would also lead to a logarithmic increase of the resistivity \cite{ALTSHULER:1980p7030}.
Both of these issues can be clarified by analyzing the MR.\\  
A uniform magnetic field breaks time-reversal symmetry and destroys the phase coherence of the ``closed path'' electronic waves, hence suppressing weak localization effects. This picture predicts a negative MR when spin effects are absent.  
In Figure \ref{Fig3}A, the MR in perpendicular field is shown for a 7 u.c. sample for different temperatures.\ The latter is found to be negative and its magnitude steadily decreases with increasing temperature, consistent with weak localization.\ The negative sign of the perpendicular field MR is evidence for the fact that the low-temperature resistivity behavior is not dominated by electron-electron interactions, as these would lead to a positive MR \cite{ALTSHULER:1980p7030}. No detectable MR was observed in metallic or insulating films.\ Figure \ref{Fig3}B shows the magnetoconductance ($\sigma(H)-\sigma(H=0))$ in units of $e^2/\pi h$ for different temperatures, $H$ being the applied magnetic field. For the weak localization regime in 2D, in the absence of spin-flip scattering, the magnetoconductance was calculated to be \cite{Lee:1985p948}
\begin{equation}\label{magnetor}
\sigma(H,T)-\sigma(H=0,T)=\frac{e^2}{\pi h}[\psi(\frac{1}{2}+\frac{1}{x})+\ln(x)]
\end{equation}
where $\psi$ is the digamma function and $x=l_{in}^24eH/\hbar.$
The only fitting parameter here is the inelastic scattering length $l_{in}$. This theory was successfully used to fit the experimental data, as shown in Figure \ref{Fig3}B.\ In the 3D case, the magnetoconductance would behave as $H^{-1/2}$ \cite{KAWABATA:1980p6857}.\ The $T$ dependence of the inelastic scattering length, predicted to scale as $T^{-p/2}$, was determined for three different samples (see Figure \ref{Fig3}C) leading to a $l_{in}\sim T^{-1/2}$ variation. This form confirms that $p=1$, as was previously deduced from resistance versus temperature measurements. We also note that $l_{in}$ is about 10 nm at 6 K, which is about four times larger than the film thickness, hence indicating that these films are indeed in the 2D regime.\\
In order to further confirm the 2D weak localization behavior, we also measured the MR for magnetic fields applied parallel to the film.
In a 2D system, and in absence of spin-orbit scatterers or magnetic impurities, orbital contributions to the MR are quenched in parallel fields. Figure \ref{Fig4} shows the in- and out-of-plane MR measured at 1.5 K for a 7 u.c. LNO film. Remarkably, for both field orientations, the MR is the same. Furthermore, an angular sweep of the magnetic field from the perpendicular to parallel configuration exhibited no particular features (resistance variations of less than 0.01 $\%$), indicating a nearly isotropic MR. This surprising result can in fact be understood within the Maekawa-Fukuyama theory \cite{Maekawa}.
It can be shown that, in the presence of significant spin fluctuations, the above two characteristics of the MR  emerge naturally when one includes magnetic terms in the diffusive electronic scattering processes. In that regime,  Eq.(\ref{magnetor}) can still be used to fit the MR, provided we define ${1\over{\tau_{in}}}= {1\over{\tau_{\varepsilon}}}+{2\over{\tau_{s}}}$ (${l^2}_{in}=D\tau_{in}$, where $D$ is the diffusion constant). The energy (magnetic) relaxation times $\tau_{\varepsilon}$ ($\tau_{s}$) are such that $\tau_{s}>\tau_{\varepsilon}$ and ${{\tau_{\varepsilon}}\over{\tau_{s}}}>\beta \frac{D^2 e^2}{\mu^2_B}$. $\beta$ is a coefficient on the order of $0.06$ in the low field limit.\\
One possible origin of these spin fluctuations could be the slight oxygen deficiency in the LNO films that leads to the formation of magnetic Ni$^{2+}$ ions. 
We may surmise that, owing to the ultrathin character of the films,  the concentration of vacancies may be large enough to preclude the formation of Kondo singlets in the metallic phase of the film. 
For this conducting state, interactions between spins of the Ni$^{2+}$ atoms are expected to be of the RKKY type such that, in the presence of disorder --recall that the system is close to a MI transition-- they may promote a spin-glass type of ordering. We note that in bulk LNO samples,  for  large enough densities of vacancies, spin glass behavior has indeed been reported \cite{Sanchez:1996p56,Gayathri:1998p1857}. In 2D, a true spin-glass transition would only take place at $T=0$ K \cite{Binder:1986}, and one does not expect the kind of static and time dependent anomalies that are the hallmark of this state. Yet, if T is low enough, one anticipates a fair amount of magnetic fluctuations and scattering on these fluctuations by diffusing electrons is expected to dominate orbital contributions and to impact the MR as discussed above \cite{Nigam:1983}.\\
Another possible origin of the magnetic fluctuations is the proximity to a charge ordered antiferromagnetic insulating state characteristic of RNiO$_3$ with the rare earth R$\neq$La \cite{medarde1997}. Recent work has demonstrated that unlike bulk LNO, ultrathin layers in a superlattice configuration display ordering phenomena such as charge and orbital ordering as well as possible antiferromagnetic ordering \cite{Liu2011,Benckiser2011,Gibert2011}. 

In summary, transport measurements in ultrathin LNO films revealed an evolution from a metallic to a strongly insulating behavior, as the room temperature value of the sheet resistance approaches $h/e^2=25\, \rm{k}\Omega$. In the intermediate regime, a transition region was found where weak localization is observed.\ The MI transition can  be associated with a dimensional crossover from 3D to 2D, which emphasizes the role of disorder. A detailed investigation of the transport properties in the weak localization regime does not reveal any obvious signatures of strong electron-electron interactions. However, such interactions cannot be ruled out in the thinner, strongly localized films, where correlations may become more important. The unexpected observation of isotropic MR also reveals the presence of magnetic scattering, suggesting the proximity to either a spin-glass state or a charge ordered antiferromagnetic state akin that of the other members of the RNiO$_3$ family.\\
We thank M. Medarde, P. Aebi, M. Garcia-Fernandez, A.D. Caviglia, A. F$\hat{\rm{e}}$te, G. Seyfarth and D. Jaccard for helpful discussions. This work was supported by the Swiss National Science Foundation through the National Center of Competence in Research, Materials with Novel Electronic Properties, MaNEP and Division II.

\bibliographystyle{revtex}

\end{document}